\documentclass[twocolumn,showpacs,superscriptaddress,aps]{revtex4}
\usepackage{graphicx}
\usepackage{amsmath}
\usepackage{amsthm}
\usepackage{amssymb}
\usepackage{latexsym}
\usepackage{array}
\usepackage{float}
\usepackage{amsfonts}
\usepackage{mathrsfs}
\usepackage{verbatim}
\usepackage{color}

\begin{document}

\title{Generalized non-locality criteria under the correlation symmetry}
\author{Kwangil Bae}
\affiliation{Sogang University, Mapo-gu, Sinsu-dong, Seoul 121-742, Korea}
\author{Wonmin Son}
\affiliation{Sogang University, Mapo-gu, Sinsu-dong, Seoul 121-742, Korea}

\received{\today}

\begin{abstract}
The most general class of non-locality criteria for $N$-partite $d$-chotomic systems with $k$ number of measurement settings is derived under the constraint of measurement symmetries. It is the complete characterisation of the multi-partite non-locality when the correlation is assumed to be symmetric under the choice of measurement settings. The generalized non-locality condition is obtained using the correlation functions, which are derived from Fourier analysis of probability spectrums. It is found that the condition for the local hidden variable (LHV) model is violated by multipartite quantum states and general constraints for the quantum violation of maximally entangled state has been obtained.
\end{abstract}


\maketitle

\newcommand{\bra}[1]{\left<#1\right|}
\newcommand{\ket}[1]{\left|#1\right>}
\newcommand{\abs}[1]{\left|#1\right|}
\newcommand{\expt}[1]{\left<#1\right>}
\newcommand{\braket}[2]{\left<{#1}|{#2}\right>}
\newcommand{\commt}[2]{\left[{#1},{#2}\right]}
\newcommand{\tr}[1]{\mbox{Tr}{#1}}

\section{Introduction}

Since John Bell formulated a condition for local realistic model in bi-partite two-level systems \cite{Bell64}, generalization of the theorem to an arbitrary large quantum system, with many measurement settings, become one of the most challenging topics in the study of quantum information science \cite{Clauser69, Mermin90,Peres99, Werner01a, Zukowski02,Collins02, Seevinck02, Barrett06, Son06,Deng09,Grandjean12,Epping13}. The problem is closely related to the possible characterization of large  quantum system \cite{Mermin90}, quantum key distribution scheme \cite{Barrett06}, network characterization \cite{Cabello14} and entanglement detection in many-body systems \cite{Werner01,Vertesi14,Tura14}.

Initially, the generalization of local hidden variable (LHV) model to an arbitrary large number of system was formulated through the term ``all the Bell inequalities'' by A. Peres about two decades ago \cite{Peres99}. Although the problem has been challenged by various methods, the full solution of the problem is not yet obtained till now \cite{Werner01a, Zukowski02}. The problem is highly non-trivial and is classified as a NP-hard problem. The complexity is originated from the fact that the formulation entails the exponentially large parameter space with respect to the probability events. Even though it is onerous investigation, the full generalisation is quite important as, for instance, it will provide an important vehicle to investigate the network of quantum correlations in a macroscopic system as well as the LHV models of a complex system.

Generalized Bell-type inequality for many-particle system provides the important benchmark for the non-trivial correlation among multipartite quantum states. In order to quantify more general type of multiparty entanglement, family of the inequalities for multipartite systems with $d$ measurement outcome has been studied many times and interesting class of inequalities are obtained through the various investigations \cite{Deng09}. Main idea in the most of the approaches is originated from Svetlichny's $N$-separability condition \cite{Seevinck02}. They obtained a set of Bell inequalities up to the limited dimension by quantifying the multiparty correlation \cite{Grandjean12}. In spite of the progress, full generalization of the multipartite Bell function has not been made yet and it is unclear whether the generalization is possible through the conventional numerical approaches using the probability polytope.

The experiment for the non-locality test with many measurement settings is also important direction for the Bell test generalization. There were many known cases that the increasing number of measurements can identify non-trivial entangled states, otherwise not being possible with two measurement settings only \cite{Laskowski04}. The generalization has been made for the two-dimensional systems and the similar approaches for the case of high dimensional system is needed to be made.

Symmetries in the generalized non-locality test is less studied in the literatures so far. Although there has been a couple of efforts to express the condition of non-locality in terms of the general set of symmetric correlation, the approaches are either limited by their scenario \cite{Zukowski02, Epping13} or limited by additional assumption such as party-swapping only \cite{Bancal10}. The derivations are not extensive because the generalisation of hidden variable test has been formulated through the probability polytope without consideration of its structural redundancies in their non-local correlation. Here, we tries generic approaches for the correlation symmetry in the generalized correlation function. Such symmetries in the non-local correlation is possible to be identified from the most generic form of the correlation as
\begin{eqnarray}
\label{eq:1}
{\cal B}&=& \sum_{\bar{n}} \sum_{\bar{m}} f_{\bar{n}, \bar{m}}E_{n_1, n_2, \cdots}(m_1, m_2,\cdots)\\
&=&\sum_{\bar{m}} \sum_{\bar{\alpha}} g_{\bar{\alpha}, \bar{m}}p(\alpha_1,\alpha_2,\cdots|m_1, m_2,\cdots)
\nonumber
\end{eqnarray}
where $\bar{m}$ the shorten notation for the vector indices of measurement settings at each site $(m_1, m_2,\cdots)$. $\bar{\alpha}$ and $\bar{n}$ are their measurement outcomes $(\alpha_1,\alpha_2,\cdots)$ and the corresponding high-order correlation indices $(n_1, n_2, \cdots)$, respectively. The functions $f$ and $g$ are weights of the correlation function $E$ and probability distribution $p$ respectively. In particular, $g$ is  denoted as generation function whose relation with $f$ will be shown in the following section.
More precise definitions of the functions will also be given in the section. Depending upon the functional distribution of $g_{\bar{\alpha},\bar{m}}$ with respect to $\bar{\alpha}$ and $\bar{m}$, the symmetries in the correlation can be found as they indicate the probability weight for the total correlation. 

Recalling the simplest inequality by Clasuer {\it et. al.} \cite{Clauser69}, the correlation is defined under the symmetric constraints as it is $$E_{CHSH}(m_1, m_2)=\sum_{\bar{\alpha}} g^{CHSH}_{\bar{\alpha},\bar{m}}p(\alpha_1,\alpha_2|m_1, m_2)$$
with $g_{\bar{\alpha},\bar{m}}^{CHSH}=(-1)^{\alpha_1+\alpha_2+m_1 m_2}$ and $\alpha_i, m_i\in\{0,1\}$. In that case, the symmetry for the exchange of measurement settings can be found from the invariance of correlation under the measurement index exchange $m_1\leftrightarrow m_2$ which allows to count the same parity measurements with equal weight. Additionally, there exists party exchange symmetry that the correlation function is invariant under the outcome index swap. The invariance is caused from the fact that the combination of measured values can be arranged in an arbitrary manner as there is no preferred order of indices for the outcome sequences. In our derivation of generalized non-locality function, similar symmetric conditions in a generalized form are considered as we impose relevant constraints in the formulation of the generalized correlation function $E$.

In this paper, we provide a method for the analytic construction of the most general correlation for the non-locality test of high-dimensional systems under the symmetric constraint. We also found maximum upper bounds of the correlation for the local hidden variable model as it is discussed in the Section II. Subsequently, it is demonstrated that the well-known inequalities with symmetric condition for the multipartite non-locality can be derived from the generic form of the correlation. The examples are given in the section III. Furthermore, we also provide the condition that the inequalities are violated by maximally entangled states as the bound can be claimed to be the criteria for the local realistic model. The condition of the quantum violation by maximally entangled state has been discussed in section IV. The technical details for the formulation are presented in the appendix, section IV, following after the main text.

\section{General class of Bell's inequality}
The class of Bell's inequalities is determined by the three parameters $N$, $k$ and $d$; $N$ local parties measure their systems with $k$ possible choices of observables which result in $d$ different measurement outcomes respectively. Once $N$, $k$ and $d$ are dertermined, we can write conditional probabilities which constitute full information about the system \cite{Schrodinger35}: a set of functions, which constitute a catalogue of full information, take the form of conditional probabilities
$p(\alpha_1,\alpha_2,\cdots, \alpha_N|m_1,m_2,\cdots,m_N)$
with $1\le \alpha_j\le d$ and $0\le m_j\le k-1$, where $\alpha_j$ is an integer number indicating the measurement outcome index for a particular choice of the observable $m_j$. A Bell correlation function, which is experimentally measurable, is derived from the one-to-one correspondence between the probabilities and multi-partite high-order correlation functions. With these functions, we can then formulate generalized Bell inequalities as follows.

Mapping the measurement outcomes to $d$ different values, the most general correlation function can be written as
\begin{equation}
\label{eq:correl}
E_{n_1,n_2,\cdots}(m_1,m_2,\cdots)=\left\langle \prod\limits_{j=1}^N  A_j^{n_j}(m_j)\right\rangle
\end{equation}
where $1\le j\le N$ is the site index. We define a shorthand notation $E_{n_1,n_2,\cdots}(m_1,m_2,\cdots):=E_{\vec{n}}(\vec{m})$. By mapping the measurement outcomes $A_j(m_j)$ to one of the complex values among the $d$ root of unity, $A_j(m_j)=\omega^{\alpha_j(m_j)}$, the correlation function takes a complex valued number with $\omega=\exp(i2\pi/d)$ and $\alpha_j\in\{1,2,\cdots, d\}$. The function {\it allowing predetermined values of a measurement}, is evaluated by the expectation of the measured values (outcomes)
\begin{equation}
E_{\vec{n}}(\vec{m})=\sum_{\{\alpha_j\}=1}^d \omega^{\vec{n}\cdot\vec{\alpha}} p(\vec{\alpha}|\vec{m})
\label{eq:corr}
\end{equation}
where $\vec{\alpha}:=(\alpha_1(m_1),\alpha_2(m_2),\cdots,\alpha_N(m_k))$. $\alpha_j(m_j)$ is determined by the physical process of the measurement whose full structure is {\it hidden} and they are usually encapsulated in the form of {\it abstracted variables}, conventionally denoted by $\lambda$. Here, we consider the case that the number of outcomes is symmetric at each party although an equivalent mapping is also possible for the case of asymmetric measurement outcomes. If the dimensions of each measurement are asymmetric, the outcome ranges at each party become different as $1\le\alpha_j\le d_j$ as thus for $\omega_j=\exp{(2\pi i/d_j)}$. Unless stated otherwise, our discussion is limited to a symmetric case while the generalization to the case of asymmetric measurement settings should be straightforward.

It is notable that the correlation obtained from the probabilities can be defined differently in general. That is because the combination of measured values can be arranged in an arbitrary manner as there is no preferred order of indices for the outcome sequences. In order to consider the full sequences of arbitrary combinations, it is necessary to introduce extra integer index $c$ for a type of correlation:
$E_{\vec{n}_c}(\vec{m})=\sum_{\{\alpha_j\}=1}^d \omega^{\vec{n}_c\cdot\vec{\alpha}} p(\vec{\alpha}|\vec{m})$
where $\vec{n}_c\equiv(c_1n_1, c_2n_2,c_3n_3,\cdots)$ with $c_j\in\{1,-1\}$. The subscript $c$ takes an integer value between 1 and $2^N$ as it specifies the type of possible correlations. Subsequently, it is possible to recover all the spectra of relevant probabilities $p(\vec{\alpha}|\vec{m})=\frac{1}{d^N}\sum_{\{n_j\}=1}^d \omega^{-\vec{n}_c\cdot\vec{\alpha}} E_{\vec{n}_c}(\vec{m})$ where the total number of distinguishable probabilities are $d^Nk^N$. The equality provides the one-to-one correspondence between probabilities for measurement outcomes and the high-order correlations.

Without any other constraints, the probability function satisfies (i) positivity $p(\vec{\alpha}|\vec{m})\ge 0, ~\forall \alpha_j, \forall m_j$ and (ii) the normalization condition $\sum_{\{\alpha_j\}}p(\vec{\alpha}|\vec{m})=1$ as the correlation function is required to satisfy $|E_{\vec{n}_c}(\vec{m})|\le  1$ and $E_{\vec{d}_c}(\vec{m})=1$ for any choices of $\vec{m}$ with $\vec{d}_c=(c_1 d, c_2 d,\cdots)$. Differently from the previous approaches in the complex valued observable \cite{Son06, Arnault12}, the complete set of distintive correlations has been identified with the parity factors $\{c_j\}$ as it generates all the possible distintive index matching to the measurement outcomes. The set of all the possible multiparty correlations then allow us to derive the Bell inequalities of the most general class including an arbitrary number of measurement settings.

If we generalize the interdependency of the probabilities from different measurements, due to incompatibilities, it is possible to derive a Bell-type inequality in the most general situation. This is the main result of this work. Using the definition, the most general Bell function $G_{N,k,d}^c$ is given by
\begin{widetext}
\begin{eqnarray}
\label{eq:GBell}
G_{N,k,d}^c:=\sum _{\{\alpha_j\}=1}^d\sum_{\{m_j\}=0}^{k-1} g_{\vec{\alpha},\vec{m}}^c p(\vec{\alpha}|\vec{m})
=\sum_{\vec{n}=1}^{d-1}
f(\vec{n}_c)\left\langle\prod\limits_{j=1}^N \left[\sum_{m_j=0}^{k-1} \omega^{c_j n_j m_j /k}A_j^{c_j n_j}(\lambda, m_j)\right]\right\rangle+\mbox{c.c.}
\label{eq:GBell1}
\end{eqnarray}
\end{widetext}
where $c.c.$ denotes the complex conjugate. The coefficient $g_{\vec{\alpha},\vec{m}}^c$ is a real function governing the linear sum of the probabilities and $f(\vec{n}_c)$ is a complex weighting function for the correlation. For the LHV model, the average is taken over the hidden variable $\lambda$ for the measurement which determines the value of the measurement function $A_j$.

The choice of the function $f(\vec{n}_c)\equiv f(c_1n_1, c_2n_2, \cdots)$ specifies the weight of each high-order correlation function in the sum and is related to the coefficient $g_{\vec{\alpha},\vec{m}}^c$ as
\begin{equation}
\label{eq:gfunction}
g_{\vec{\alpha},\vec{m}}^c= 2 ~\mbox{Re}\left[\sum_{\vec{n}}f(\vec{n}_c)\omega^{\vec{n}_c\cdot(\vec{\alpha}+\vec{m}/k)}\right].
\end{equation}
 From its Fourier analysis, it is also possible to show that the choice of function $f(\vec{n}_c)$ generates all the possible combinations of the correlations $E_{\vec{n}_c}(\vec{m})$. The decomposition of the generation function is obtained when the full correlations have been taken into account. Mathematically, it means that $\sum_{m_j=0}^{k-1} \omega^{c_j n_j m_j /k}A_j^{c_j n_j}(\lambda, m_j)=0$ when $n_j=d$, $\forall j$.

The decomposition of the correlation into the probabilities in Eq.(\ref{eq:GBell}) provides the most general correlation as it constitutes ``all the Bell inequalities'' \cite{Peres99}. The only difference in this construction from the original formulation is the condition of homogeneity and the symmetries in the choice of the measurements. It means that the number of measurements and the dimension of each party are chosen to be identical. Additionally, it also means that all the probabilities of the measurement choices are all equal as the choices are symmetrically distributed as to be completely random. In the formulation, the generalized correlation displays two symmetries. (i) Symmetric distribution of measurement: The weights of $m_i$-th and $m_j$-th measurement of party $i$ and $j$ are same when $m_j=m_i$. (ii) Symmetric under party-swapping: When the order of correlation terms is homogeneous, $n_j=n\, \forall\, n_j$, the Bell function is invariant under any permutation of party index $j$.

It can be shown that all the known Bell functions within the homogeneous condition can be derived as a special case of the function $G$ in (\ref{eq:GBell1}).

For the LHV constraint, the first decomposition in (\ref{eq:GBell}) using the probabilities is directly linked to the modified version of Farkas lemma \cite{Rockafellar70}. \\
In the formulation, it is straightforward that the local realistic (LR) bound can be obtained by the Farkas vector $g_{\vec{\alpha},\vec{m}}^c$ after it is optimized over the all the measurement outcomes as $\max_{\vec{\alpha}}\Big[\sum_{\vec{m}} g_{\vec{\alpha},\vec{m}}^c\Big]$\cite{Sup}. The application of the Farkas lemma is explained more detailed in the appendix, Section \ref{sec:local}. The bound for the correlation $G_{N,k,d}^c$ is obtained as
\begin{eqnarray}
\label{eq:LRB}
G_{N,k,d}^c&\le& B_{LR}=\max_{\vec{\alpha}}\Big[\sum_{\vec{m}} g_{\vec{\alpha},\vec{m}}^c\Big]
\end{eqnarray}
which {\it provides the most general criteria for the probabilities allowed by the LHV model}. Evaluation of the bound requires the functional optimization over the measured values, and the analytic evaluation is possible. It can be achieved by the specification of local parameters under the functional constraints. The analytical values can be obtained efficiently if one follows the optimal counting method described explicitly as it is illustrated in the previous work of us \cite{Bae14}. The usefulness of our formalism in the calculation of local bound (\ref{eq:LRB}) is discussed in the appendix section with examples.

The decomposition of the correlation function for possible local measurements under the LHV model is nonetheless trivial. It allows quantum characterization of the correlation when the coefficients $g_{\vec{\alpha},\vec{m}}^c$ as well as $f(\vec{n}_c)$ are appropriately determined. In the Bell correlation function (\ref{eq:GBell1}), the general decomposition of measurements with the weighting factor $f(\vec{n}_c)$ is obtained. In the following, we demonstrate the derivation of known Bell functions through the specification of $g_{\vec{\alpha},\vec{m}}^c$ and $f(\vec{n}_c)$. The condition for quantum violation will follow the analysis.

\section{Derivation of various inequalities from the general form}\label{sec:derivation}
First of all, we show that the function (\ref{eq:GBell}) reduces to the Clauser-Horne-Shimony-Holt (CHSH)-Bell inequality \cite{Clauser69}. By taking the generation function $g_{\vec{\alpha},\vec{m}}^{CHSH}=(-1)^{\alpha_1+\alpha_2+m_1 m_2}$, it is straightforward to show that the left hand side becomes CHSH inequality. Through the evaluation of Eq. (\ref{eq:LRB}), it is possible to obtain
 $$G_{2,2,2}^{CHSH}\le 2, $$
as it is known as standard Bell-CHSH inequality. Here, one can find the Fourier transformed the generating function which is given as $f_{n_1,n_2}^{CHSH}=(1- i)\delta_{n_1,1}\delta_{n_2,1}/2$.

The correlation function becomes Collins-Gisin-Linden-Massar-Popescu (CGLMP) function \cite{Collins02} for (2,2,d)-class system when the generating function takes the form
\begin{eqnarray}
g_{\vec{\alpha}, \vec{m}}^{CGLMP}&=&(-1)^{m_1-m_2}\sum_{k=0}^{d-1}\left(1-\frac{2k}{d-1}\right) \\
&~&~~~~~~~~~\times\delta^d(\alpha_2-\alpha_1-k-z(m_1 m_2))\nonumber
\end{eqnarray}
where $\delta^d$ is kroneker delta function in the modulo $d$ space and $z(m_1, m_2)$ is a binary function mapping $z(0,0)=z(1,1)=z(1,0)=0$ and $z(0,1)=1$. It results in $G_{2,2,d}^{CGMLP}\le \max_{\vec{\alpha}}\Big[\sum_{\vec{m}} g_{\vec{\alpha},\vec{m}}^{CGMLP}\Big]=2$. In that case, the correlation weighting function is given by
\begin{equation}
f^{CGMLP}_{n_1, n_2}=\frac{1/2}{d-1}\sum_{n=1}^{d-1}\sec\left[\frac{n\pi}{2 d}\right]\omega^{\frac{n}{4}}\delta_{n_1=n}^d\delta_{n_2=-n}^d
\end{equation}
whose detailed derivation is shown in \cite{Sup}. The quantum violation of the LR bound can be found in the original works \cite{Clauser69, Collins02} and can be recovered in the framework of generalized correlation formalism in an analytical manner.

It can be demonstrated that the generic formula can be used to derive the Bell functions for multipartite systems such as Mermin and Zukowski-Brukner (ZB) functions \cite{Mermin90, Zukowski02}.

The Mermin function $G_{N,2,2}^{M}$ is obtained by assigning $g_{\vec{\alpha},\vec{m}}^M=(-1)^{\sum_j\alpha_j}\mbox{Re}[(i)^{\sum_j m_j }]$ when we have $f^M(1)=1/2$. From the form of $g_{\vec{\alpha},\vec{m}}$, the Mermin bound can be found as $G_{N,2,2}^{M}\le 2^{(N-1)/2}$ for odd $N$ and  $G_{N,2,2}^{M}\le 2^{N/2}$ for even $N$. Ultimately, the most general (N,2,2)-class ZB correlation can be obtained when $f^{ZB}(\vec{n})=(1-i)/2$ as
 \begin{equation}
G_{N,2,2}^{c, ZB}=\sum_{\{m_j\}=0}^1 P(m_1,m_2,\cdots)  c_1^{m_1} c_2^{m_2}\cdots E(m_1, m_2, \cdots)
\end{equation}
where the parity function $P(m_1,m_2,\cdots)\in\{-1,1\}$ takes its value $1$ for $[(\sum_j m_j) \mod 4] \in\{0, 1\}$ and $-1$ for $[(\sum_j m_j) \mod 4] \in\{2,  3\}$. The LR bound of the function can be found as $G_{N,2,2}^{c, ZB}\le 2^N$ from the probability coefficient
$$g_{\vec{\alpha},\vec{m}}^{ZB}=(-1)^{\sum_j c_j\alpha_j} \mbox{Re}[(1-i)(i)^{\sum_j c_jm_j}]$$ through the maximization of $\sum_{\vec{m}} g_{\vec{m},\vec{\alpha}}$ over a proper choice of $(\alpha_1, \alpha_2, \cdots)$ following its derivation in Eq.(\ref{eq:gfunction}). A more general constraint of the Bell theorem for the class of full correlation can be obtained as one combines $G_{N,2,2}^{c, ZB}$ for all the possible $\vec{c}$ as $\sum_{c=1}^{2^N}|G_{N,2,2}^{c, ZB}|\le 2^N$.

 For the case of $(2, k, 2)$-class Bell test scenario, the most general correlation function has been inspected by Epping {\it et. al.} \cite{Epping13}. In that case, the correlation function can take the form $G_{2,k,2}^{EKB}=\sum_{m_1, m_2} \beta_{m_1, m_2} E(m_1, m_2)$ where $E(m_1, m_2)$ is the first-order correlation function defined in Eq. (\ref{eq:correl}). The correlation function can be found when $\beta_{m_1,m_2}\equiv f(1)\omega^{(m_1+m_2)/k}+c.c.$ such that the coefficient $\beta$ can be related to the probability coefficient $g^{EKB}_{\vec{\alpha}, \vec{m}}$ as $g^{EKB}_{\vec{\alpha}, \vec{m}}=(-1)^{(\alpha_1+\alpha_2)}\beta_{m_1, m_2}$. Thus, the LR bound of the function is obtained as
\begin{eqnarray}
G_{2,k,2}^{EKB}&\le&\max_{\vec{\alpha}}\left[\sum_{\vec{m}} (-1)^{(\alpha_1+\alpha_2)}\beta_{m_1, m_2}\right]\\
&=&2|f(1)|\max_{\vec{\alpha}}\left[\sum_{\vec{m}} \cos(\pi\bar{\alpha})\cos\left(\theta_f+\frac{\pi\bar{m}}{k}\right)\right]\nonumber\\
&=& 2|f(1)|\cos\theta_f/\sin^2(\pi/2k),~~\mbox{for}~ 0\le\theta_f\le\pi/2k\nonumber
\end{eqnarray}
where $\bar{\alpha}=\alpha_1+\alpha_2$, $\bar{m}=m_1+m_2$ and $\theta_f$ is the phase factor of the complex function $f(1)$ as $f(1)=|f(1)|e^{i\theta_f}$. The second equation in the formula is obtained after the optimization through the counting the largest cosine terms. Quantum maximum of $G_{2,k,2}^{EKB, Q}$ is given by the singular value decomposition of the $\beta$ matrix, $k||\beta||_2=k^2|f(1)|$, as argued in \cite{Epping13}. Furthermore, it can be proved that the violation is optimal compared to the one with a different number of measurement settings at each site as $k_1$ and $k_2$ and it can be argued that the symmetric Bell function is more resilient to experimental noise and inefficiencies than the asymmetric case.

\section{Condition of quantum violation by maximally entangled state}
For quantum systems, the measured variables are expressed by eigenvalues of an operator whose expectation corresponds to the statistical average of measurement values. In that case, the decomposition in Eq.(\ref{eq:GBell}) can be represented by measurement operators whose explicit form is given by
\begin{equation}
\hat{A}_j(m_j)=\sum_{\alpha} \omega^{\alpha}|A_{\alpha}(m_j)\rangle\langle A_{\alpha}(m_j)|
\end{equation}
where $d$-dimensional orthogonal bases $|A_{\alpha}(m_j)\rangle$ are chosen to satisfy $\langle A_{\alpha}(m_j)|A_{\beta}(m_j)\rangle=\delta_{\alpha, \beta}$.
The bases $|A_{\alpha}(m_j)\rangle$ can be obtained as a linear combination of the orthogonal computational bases, $|A_{\alpha}(m_j)\rangle=\frac{1}{\sqrt{d}}\sum_{\beta=1}^{d} \omega^{\beta(\alpha+m_j/k)}|\beta\rangle$ where the $m_j$-th basis has been obtained by the phase shift of the fourier transformed state whose phase shift is distributed from 0 to $k-1$ evenly. In the sense that the measurement bases are evenly distributed in Hilbert space, the bases can constitute a maximal test.

From the bases, one can show that the spectral sum of measurement operators become ladder lowering operators as
\begin{equation}
\sum_{m_j=0}^{k-1} \omega^{ n_j m_j /k}\hat{A}_j^{n_j}( m_j)=k \sum_{\beta}|\beta\rangle\langle\beta+n_j| \equiv k \hat{J_j}^{n_j}
\end{equation}
which corresponds to the $n_j$-th power of a lowering operator $\hat{J_j}=\sum_{\beta}|\beta\rangle_j\langle\beta+1|$ for a high dimensional state as $(\hat{J_j})^{n_j}|\alpha\rangle=|\alpha-n_j\rangle$. In addition, a phase shift operator $\hat{P}_{\nu}$ acting on the orthogonal computational bases generates an extra phase $\hat{P}_{\nu}|\alpha\rangle=\omega^{-\nu\alpha}|\alpha\rangle$ and can be used for the local unitary transform on the lowering operator as $\hat{P}_{\nu}^{\dagger}\hat{J}_j^{n_j}\hat{P}_{\nu}=\omega^{\nu n_j}\hat{J}_j^{n_j}$. The phase shift operation is effective in order to obtain the different measurement bases that produce the correlation values beyond LR bounds.

Together with the sequence of the local phase shift operation $\hat{P}_{\nu_j}$, the generalized Bell function for a quantum state can be obtained in terms of high-order correlation functions. If there is a state whose expectation goes beyond the LR bound, the state cannot be described by the LHV model. With the measurements and the local rotations, the correlation function of a quantum state becomes
\begin{eqnarray}
G_{N,k,d}^Q= k^N \sum_{\vec{n}=1}^{d-1}
f(\vec{n})\omega^{\vec{\nu}\cdot\vec{n}}\left\langle\overset{N}{\underset{j=1}{\bigotimes}} \hat{J_j}^{n_j}\right\rangle+\mbox{c.c.}
\label{eq:qcorr}
\end{eqnarray}
where $\vec{\nu}=(\nu_1, \nu_2, \cdots)$ represents the composite components of local phase shifts at each site.

In the following, we show that the LR bound is violated by a simple symmetric quantum state with an appropriate choice of $f(\vec{n})$. When the powers of the lowering operators at each site are uniform, $n_1=n_2=\cdots=n$, the $N$-partite maximally entangled (ME) pure quantum state $|\psi\rangle=\sum_{\alpha}1/\sqrt{d}~|\alpha\rangle^{\otimes N}$ gives the quantum correlation $G_{N,k,d}^{Q,ME}= k^N \sum_{n=1}^{d-1}\left(1-\frac{n}{d}\right)f(n)\omega^{(n\sum_j\nu_j)}+c.c.$ with an arbitrary local parameter $\nu_j$ representing a choice of measurements. If $\vec{\nu}$ is chosen to satisfy that $-n\sum_j\nu_j=\mbox{Arg}[f(n)]$, the value of the quantum correlation is upper bounded by the quantum maximum for the ME state $G_{N,k,d}^{Q,ME}\le 2 k^N \sum_{n=1}^{d-1}\left(1-\frac{n}{d}\right)\left|f(n)\right|\equiv Q_M$. Thus, an appropriate specification of $f(n)$ results in the violation of LR bound
\begin{equation}
B_{LR} \le Q_M=2 k^N \sum_{n=1}^{d-1}\left(1-\frac{n}{d}\right)\left|f(n)\right|
\end{equation}
the bound $B_{LR}$ is also determined by the function $f(n)$, as shown in (\ref{eq:LRB}). Therefore, the general constraint for $B_{LR} < Q_M$ can be obtained from the appropriate choice of the weighting factors. If $|f(n)|$ is either a constant function or a monotonically increasing function with respect to $n/d$, then a general trend of violation, $B_{LR} < Q_M$, can be obtained. Generally speaking, the values of $B_{LR}$ determined from $f(\vec{n})$ provide the criteria for the LHV model and constitutes the generalized Bell function as long as $B_{LR} < Q_M$. Explicit criteria in a couple of special cases and their analysis can be found in \cite{Sup}.

The quantum upper bound of CGMLP equation for the ME state can be analytically formulated. For the ME state, the quantum expection of general Bell function becomes
\begin{equation}
2 < Q^{CGMLP}_M=\frac{4}{d-1}\sum_{n=1}^{d-1}\left(1-\frac{n}{d}\right)\sec\left[\frac{n\pi}{2d}\right]
\end{equation}
for any value of $d$. Thus, the state is non-local although the value is not quantum maximum. It is known that the maximal violation of the CGMLP inequality is obtained by partially entangled state. In order to obtain the quantum maximum, the correlation (\ref{eq:qcorr}) should be evaluated for a partially entangled quantum state and  be optimized by the parameters of the state. In that way, the maximum value of violation can also be derived from our formalism as it is illustrated in \cite{Sup}.

\section{Remarks}
In this work, we derived general criteria of the LHV model analytically and provided examples that violate the condition. The derivation was done through the one-to-one correspondence between the general probability space and the correlations of arbitrary high-order moments. We proved that the decomposition of the correlation function using joint probabilities gives the straightforward LR bound. The decomposition by the full correlations of high-order moments can be used for non-trivial quantum violation under the provided settings of measurements. The result sheds the light on the general characterization of quantum correlation in an arbitrary number of high dimensional systems and the arbitrary number of measurement settings.

{\it Acknowledgments} -
The author acknowledge G. Bae and M.S. Kim for their useful discussions. 

\section{Appendix}
In this appendix, we present that (i) the explicit derivation of the equivalence between the decompositions for the generalized Bell function using the full measurement probabilities and the complete set of distintive multipartite correlations, (ii) the derivation of several major known Bell functions from the generalized formalism and (iii) the method to obtain the local realistic bounds. They provide the explicit derivations of the formulae in the main article, in order to support the argument that the decomposition of our multipartite correlations leads to the most general class of Bell's inequalities.
\subsection{Equivalent form of correlation}
Equivalance between the probability polytope and the convex sum of high order multipartite correlation can be provided from its definition. Both of them can be used to construct the general multipartite correlation, however, they represent different aspects of the multipartite correlations. One is useful for an explicit form of the physical measurements in the test of the local realistic model while the other can be used for the direct quantum maximal value in a straight forward manner.
The connection between the probability and the correlation for generic Bell fucntion can be derived from their original definition. As its first step, the general Bell function can be represented by convex combination of probabilities as
\begin{widetext}
\begin{eqnarray}
\label{eq:GBell}
G_{N,k,d}^c&:=&\sum _{\{\alpha_j\}=1}^d\sum_{\{m_j\}=0}^{k-1} g_{\vec{\alpha},\vec{m}}^c p(\vec{\alpha}|\vec{m})=\sum _{\{\alpha_j\}=1}^d\sum_{\{m_j\}=0}^{k-1}\frac{1}{2}\left[ F_{\vec{\alpha},\vec{m}}^c+{F_{\vec{\alpha},\vec{m}}^c}^* \right]p(\vec{\alpha}|\vec{m})
\end{eqnarray}
where we decompose the coefficient of the probability in the correlation, $g_{\vec{\alpha},\vec{m}}^c$,  into an arbitrary imaginary function $F_{\vec{\alpha},\vec{m}}^c$. In fact, it allows us to obtain the Bell function using the high order correlation functions as
\begin{eqnarray}
G_{N,k,d}^c&=&\frac{1}{2 d^N}\sum_{\{n_j\}=1}^d\sum_{\{m_j\}=0}^{k-1} \sum _{\{\alpha_j\}=1}^dF_{\vec{\alpha},\vec{m}}^c  \omega^{-\vec{n}_c\cdot\vec{\alpha}} E_{\vec{n}_c}(\vec{m}) +c.c.
\nonumber\\
&=&\frac{1}{2 d^N}\sum_{\{n_j\}=1}^d\sum_{\{m_j\}=0}^{k-1} \sum _{\{\alpha_j\}=1}^d F_{\vec{\alpha},\vec{m}}^c \omega^{-\vec{n}_c\cdot\vec{\alpha}}  \left\langle \prod\limits_{j=1}^N  A_j^{c_j n_j}(m_j)\right\rangle_{avg}+c.c.
\\
&=&\sum_{\vec{n}=1}^{d-1}
f(\vec{n}_c)\left\langle\prod\limits_{j=1}^N \left[\sum_{m_j=0}^{k-1} \omega^{c_j  n_j m_j /k}A_j^{n_j}(\lambda, m_j)\right]\right\rangle+\mbox{c.c.}\nonumber
\label{eq:GBell1}
\end{eqnarray}
\end{widetext}
where we define
\begin{equation}
\frac{1}{2 d^N}\sum _{\{\alpha_j\}=1}^d F_{\vec{\alpha},\vec{m}}^c\omega^{-\vec{n}_c\cdot\vec{\alpha}}= f(\vec{n}_c) \omega^{\vec{n}_c \cdot\vec{m} /k}.
\end{equation}
From its definition, the function $F_{\vec{\alpha},\vec{m}}^c$ can be obtained through the fourier transformation of $f(\vec{n}_c)$ and thus we have
\begin{equation}
\label{eq:gfunction}
g_{\vec{\alpha},\vec{m}}^c= 2 ~\mbox{Re}\left[\sum_{\vec{n}}
f(\vec{n}_c)\omega^{\vec{n}_c\cdot(\vec{\alpha}+\vec{m}/k)}\right]
\end{equation}
that relates the function $f$ for the high-order correlations and $g$ for the coefficient strength of the measurement probabilities. From the equivalence, it can be identified that the Bell function can be expressed by convex sum of probabilities and it can be decomposed in terms of correlation functions. It also means that the problem of the generalized Bell inequality can be addressed in terms of convex set of general probabilities as well as correlation functions of multiparty systems in an equivalent manner.

\subsection{Local realistic upper bound}\label{sec:local}
The structural construction of Bell's inequality is given as following. For the case of Bell function, it can take a value of the algebraic upper bound when the system is subjected to a local realistic model. Statement of the model is that the expectation value of a correlation under a local choice of measurements $a$ and $b$ can be obtained as
\begin{equation}
\langle A B\rangle=\sum_{\lambda} \rho(\lambda) A(a, \lambda) B(b,\lambda)
\end{equation}
where $\rho(\lambda)$ is a probability density as a function of a hidden variable $\lambda$. Here, we assumed that the variable is distributed in a discrete manner.

Validity of the model is also provided by the existance of legitimate probabilities  $\rho(\lambda)$ for the hidden variable $\lambda$ which satisfies normalization condition $\sum_{\lambda}\rho(\lambda)=1$. It means that the spectrum of probabilities $p(\alpha_1, \alpha_2| m_1, m_2)$, for all the possible outcomes from the experimental tests can be decomposed by the probabilities of a state which is determined by the action of hidden variables $\lambda$ as
\begin{equation}
p(\alpha_1, \alpha_2| m_1, m_2)=\sum_{\lambda}\rho(\lambda) B_{\lambda}^{(\alpha_1, \alpha_2| m_1, m_2)}.
\end{equation}
Here, we denote an arbitrary Boolean function,  $B_{\lambda}^{(\alpha_1, \alpha_2| m_1, m_2)}\in\{0,1\}$, in order to specify the type of a state with a specific probability distribution of hidden parameters. From the condition of the probability $p(\alpha_1, \alpha_2| m_1, m_2)$, the constraints for the Boolean function are given as
\begin{equation}
\sum_{\vec{\alpha}}B_{\lambda}^{(\vec{\alpha}| \vec{m})}=1,~~~~~
\sum_{\vec{\alpha}}\sum_{\vec{m}}B_{\lambda}^{(\vec{\alpha}| \vec{m})}=N_T
\end{equation}
where $N_T=\dim[m_1]\times\dim[m_2]$ is the total number of measurments throughout the sites. We illustrate this for bipartite system, however, it can be generalized into the arbitrary number of system in a same way.

From the constraints and the Farkas lemma \cite{Rockafellar70}, it can be shown that the local realistic upper bound of generic Bell function can be derived from the formulation of general probabilities. The statement of the lemma is that if $\sum_{\vec{\alpha},\vec{m}}g_{\vec{\alpha},\vec{m}}^c \prod_j A_j(\alpha_j|m_j,\lambda)\le B_{LR}$ is satisfied by all $\lambda$ and the local realistic (LR) bound $B_{LR}$ then $G_{N,k,d}^c=\sum_{\vec{\alpha},\vec{m}}g_{\vec{\alpha},\vec{m}}^c p(\vec{\alpha}|\vec{m})\le B_{LR}$ where $p(\vec{\alpha}|\vec{m})=\sum_{\lambda} \rho(\lambda) \prod_j A_j(\alpha_j|m_j,\lambda)$ and  $\rho(\lambda)$ is a positive function, satisfying $\sum_{\lambda} \rho(\lambda)=1$. The bound can be derived when one considers the convex sum of the probabilities,
\begin{eqnarray}
G_{N,k,d}^c&=&\sum_{\vec{\alpha}}\sum_{\vec{m}} g_{\vec{\alpha},\vec{m}}^c p(\vec{\alpha}|\vec{m})\nonumber\\
&=&\sum_{\lambda}\rho(\lambda)\left[\sum_{\vec{\alpha}}\sum_{\vec{m}} g_{\vec{\alpha},\vec{m}}^cB_{\lambda}^{(\vec{\alpha}| \vec{m})}\right]\\
&\le&\max_{\lambda}\left[\sum_{\vec{\alpha}}\sum_{\vec{m}} g_{\vec{\alpha},\vec{m}}^cB_{\lambda}^{(\vec{\alpha}| \vec{m})}\right]\le\max_{\vec{\alpha}}\Big[\sum_{\vec{m}} g_{\vec{\alpha},\vec{m}}^c\Big]\nonumber
\end{eqnarray}
where the maximal bound is found by the probability coefficient $ g_{\vec{\alpha},\vec{m}}^c$ maximized over the measurement outcomes $\vec{\alpha}$. For the inequalities, we use the fact that the convex sum of the probabilities is upper bounded by the largest coefficient in the sum, reads $c_1p_1+c_2p_2+c_3p_3+\cdots \le \max_i c_i$ when $\sum_i p_i=1$ and $0\le p_i\le1$,$\forall i$.

In this part, the usefulness of local realistic optimization presented in (\ref{eq:LRB}) is discussed with example. The optimization can be summarized as the problem of distributing optimal distribution of $\vec{\alpha}$ over $\sum_{\vec{m}} g_{\vec{\alpha},\vec{m}}^c$. Although the problem is non-trivial in general, the strategy of deriving optimal distribution of $\{\vec{\alpha} |\forall \vec{m}\}$ under consideration of the constraint on $\vec{\alpha}$ can be an useful approach to the problem. Moreover the analysis of the convexity of the generation function brings advantages in the optimization problem. \\
\indent We present the example of optimization with familiar case of CGLMP in our formalism. The original CGLMP correlation can be equivalently modified as
\begin{align}
\label{corr4}
C_d
&=\sum_{\vec{\alpha}}\sum_{\vec{m}} g_{\vec{\alpha},\vec{m}}^{CGMLP}p(\vec{\alpha}|\vec{m})\\
&=\sum_{k=0}^{d-1}\sum_{ij=0}^{1}g(k)P(\alpha_{m_1 m_2} \doteq k \mod\, d)
\end{align}
where $g(k)=1-\frac{2k}{d-1}$, $\alpha_{00}=\alpha_1(0)-\alpha_2(0)$, $\alpha_{01}=\alpha_2(1)-\alpha_1(0)$, $\alpha_{10}=\alpha_2(0)-\alpha_1(1)-1$, $\alpha_{1}=\alpha_1(1)-\alpha_2(1)$. Then the constraint $\cal{C}$ on $\vec{\alpha}$ can be expressed as $\sum_{m_1 m_2} \alpha_{m_1 m_2}= -1$. One can restricts the parameter space containing optimal distribution of outcome as $\{\vec{\alpha}  | \cal{C} \}$. The constraint is more restricted when we consider the functional convexity of $g$ and the form of local bound $B_{LR}=\max_{\vec{\alpha}}\sum_{\vec{m}} g_{\vec{\alpha},\vec{m}}^c=\max_{\{\dot{\alpha}_{m_1 m_2}\}}\sum_{m_1 m_2} g(\dot{\alpha}_{m_1 m_2})$ where $\dot{\alpha}_{m_1 m_2}$ is modulo-$d$ value of $\alpha_{m_1 m_2}$. Suppose the parameter set $\{\dot{\alpha}_{00},\dot{\alpha}_{01},\dot{\alpha}_{10},\dot{\alpha}_{11} \}$. Then one can think of the situation in which the maximal number of 0's appears in the set such that $\{0,0,0,d-1\}$. The other sets can be obtained from substituting elements in $\{0,0,0,d-1\}$ under the constraint $\cal{C}$. One way is to maintain the sum of the elements and the other way is to change the sum as the multiple of the $d-1$ larger than $d-1$. The former are achieved by adding $a$ and $-a$ to the element 0 and $d-1$ respectively. In this case correlation value is invariant as $g$ is linear i.e. $g(0)+g(d-1)=g(a)+g(d-a-1)$. The other case is given when adding arbitrary values to 0 elements such that the sum of elements result in the multiple of $d-1$. It always gives smaller correlation value because $g$ is decreasing function. Therefore the correlation is always same or smaller than $3g(0)+g(d-1)=2$ when two type of substitution is successively conducted to the set $\{0,0,0,d-1\}$. And no other case can occur. In our formalism, we always can consider functional form of the generation functions given with (\ref{eq:gfunction}). It provides the possibility of further restriction to the parameter set containing optimal case as explained above.\\
\indent Also, it might be worth noting that our formalism can be applied to the problem of the analytic derivation of facet inequalities of local ploytope in generalized Bell scenario. In our approach the number of optimal parameter set $\cal N$ corresponding to local bound can be calculated from the constraint on the outcome parameter. Deriving the constraint that maximizes $\cal N$ such that $\cal N$ is larger than the dimension of the local bound can be an analytic approach to tighten a Bell's inequality. As the condition is the necessary condition for a Bell's inequality to be tight \cite{Masanes02}.

\subsection{Derivation of correlation coefficient for CGMLP inequality}
As it has been shown, the inequalities of the general class for the characterization of the realistic model can be derived from the formalism above. The most notable class of Bell inequalities for the bipartite high dimensional system ($d\times d$ system) is the one by Collins-Gisin-Massar-Linden-Popescu (CGMLP)\cite{Collins02} and the other by Son-Lee-Kim (SLK)\cite{Son06}. While the derivation of SLK inequality from the generic correlation function is straightforward due to its original definition, the relation between CGMLP and generic Bell function is not trivial. In this section, we show how to derive CGMLP function in a different decomposition explicitly. The importance of the different decompositions lies not just in the demonstration of the generality of the generic Bell function but also in the efficient derviation of maximal bound of the (quantum as well as classical) correlation.

In its original construction \cite{Collins02}, the function for the CGMLP inequality takes the form
\begin{align}\label{Cd}
\nonumber
C_d=\sum_{k=0}^{([d/2]-1)}&\left(1-\frac{2k}{d-1}\right)\\
&\{P(A_1=B_1+k)+P(B_1=A_2+k+1)\nonumber\\
&+P(A_2=B_2+k)+P(B_2=A_1+k)\}\nonumber\\
&-\{P(A_1=B_1-k-1)+P(B_1=A_2-k)\nonumber\\
&+P(A_2=B_2-k-1)+P(B_2=A_1-k-1)\}\nonumber
\end{align}
whose local realistic upper bound is violated by a quantum state. After shuffling the probabilities which are in the equivalent classes, the distribution of probabilities can be rearranged and it can be rewritten as
\begin{eqnarray}
C_d&=&\sum_{k=0}^{d-1}\left(1-\frac{2k}{d-1}\right)\\
&&~~\times\{P(A_1=B_1+k)+P(A_2=B_2+k)\nonumber\\
&&~~~~-P(A_2=B_1+k)-P(A_1=B_2+k+1)\}\nonumber\\
&=&\sum_{\vec{\alpha}}\sum_{\vec{m}} g_{\vec{\alpha},\vec{m}}^{CGMLP}p(\vec{\alpha}|\vec{m})
\end{eqnarray}
where the coefficient takes the functional form $g_{\vec{\alpha},\vec{m}}^{CGMLP}=(-1)^{m_1-m_2}\sum_{k=0}^{d-1}\left[1-2k/(d-1)\right]\delta(\alpha_1-\alpha_2-k-z(m_1,m_2))$ with an appropriate mapping for the choice of measurements. They are indexed as $(A_1, A_2)\rightarrow (m_1=0, m_1=1)$ and $(B_1, B_2)\rightarrow (m_2=0, m_2=1)$ together with a binary function $z(m_1,m_2)\in\{0,1\}$. In this case, the binary function will take the values as $z(0,0)=z(1,1)=z(1,0)=0$ and $z(0,1)=1$. Moreover, the coefficient $g_{\vec{\alpha},\vec{m}}^{CGMLP}$ can be further decomposed as
\begin{eqnarray}
g_{\vec{\alpha},\vec{m}}^{CGMLP}
&=&\frac{(-1)^{m_1-m_2}}{d}\sum_{k=0}^{d-1}\left(1-\frac{2k}{d-1}\right)\nonumber\\
&&~~~~~~~~
\times\sum_{n=0}^{d-1}\omega^{n(\alpha_1-\alpha_2-k-z(m_1,m_2))}\\
&=&\frac{2(-1)^{m_1-m_2}}{d-1}\sum_{n=1}^{d-1}\frac{\omega^{n(\alpha_1-\alpha_2-z(m_1,m_2))}}{1-\omega^{-n}}\nonumber\\
&=&2 \mbox{Re}\Big[\sum_{n_1}\sum_{n_2} f(n_1, n_2)~ \omega^{n_1\alpha_1+n_2\alpha_2}\nonumber\\
&&~~~~~~~~~~~~~~~~~~~~~\times\omega^{(n_1m_1+n_2m_2)/2}\Big]
\end{eqnarray}
where the weighting factor $f$ is found as
\begin{equation}
f^{CGMLP}(n_1, n_2)=\frac{1/2}{d-1}\sum_{n=1}^{d-1}\sec\left[\frac{n\pi}{2 d}\right]\omega^{\frac{n}{4}}\delta_{n_1=n}\delta_{n_2=-n}.
\end{equation}
The equivalence in the last equation can be proved using the straightforward Fourier analysis with the equalities,
\begin{eqnarray}
\sum_{n=1}^{d-1}\omega^{n}&=&\sum_{n=1}^{d-1}\omega^{-n},
~~~\sum_{n=1}^{d-1}\omega^{n/2}=-\sum_{n=1}^{d-1}\omega^{-n/2},\nonumber\\
\sec\left[\frac{n\pi}{2 d}\right]&=&\frac{2\omega^{-n/4}}{1+\omega^{-n/2}}.
\end{eqnarray}
Therefore, the derivation of CGMLP correlation function is possible from the generalized correlation formalism in Eq. (\ref{eq:GBell1}). From the explicit expression of $g_{\vec{\alpha},\vec{m}}^{CGMLP}$, it is not difficult to find the local realistic bound of CGMLP function as $\max_{\vec{\alpha}}\Big[\sum_{\vec{m}} g_{\vec{\alpha},\vec{m}}^{CGMLP}\Big]=2$.

Violation of the inequality by a quantum state can be inspected further and maximum value of the correlation for a quantum state is still under investigation. With the coefficients that had been found in the previous section, a quantum correlation for CGMLP inequality can be obtained. After Schmidt decomposition, a bipartite pure state can be written in general as $|\psi\rangle=\sum_{n=0}^{d-1} \gamma_n |n, d-1-n\rangle$ and the CGMLP correlation function can be written as
\begin{eqnarray}
G_{2,2,d}^{Q,CGMLP}&=&2^2\sum_{n=1}^{d-1}f^{CGMLP}_n\sum_{\alpha=n}^{d-1}\gamma_{\alpha-n}^*\gamma_n+c.c.\nonumber\\
&=&\frac{2}{d-1}\sum_{n=1}^{d-1}\sec\left[\frac{n\pi}{2 d}\right]\omega^{\frac{n}{4}}\Omega_n+c.c.
\end{eqnarray}
where $\Omega_n\equiv\sum_{\alpha=n}^{d-1}\gamma_{\alpha-n}^*\gamma_n$. Upto the local unitary phase shift, the correlation function is upper bounded by the real component of $\Omega_n$ as
\begin{equation}
G_{2,2,d}^{Q,CGMLP}\le \frac{4}{d-1}\sum_{n=1}^{d-1}\sec\left[\frac{n\pi}{2 d}\right] \mbox{Re}[\Omega_n]
\end{equation}
where the parameters in the upper bound have a constraint $\Omega_0=1$. For example, when the system is two dimensional, $d=2$, the upper bound is characterized by a single unknown parameter $\mbox{Re}[\Omega_1]=\mbox{Re}[\gamma_0^*\gamma_1]$ with a constraint equation $\Omega_0=|\gamma_0|^2+|\gamma_1|^2=1$ and it results in the maximal bound $2\sqrt{2}$. In the same way, further generalization is possible when the optimal $\Omega_n$ for the maximum $G_{2,2,d}^{Q,CGMLP}$ is found with an appropriate parameterization of $\gamma_i$.

\subsection{N number of two outcome systems with two measurements at each site}
For the case of Mermin inequality, testing the (N,2,2)-class system, the measurement function is defined,
\begin{equation}
M=\frac{1}{2}\left[ (\sigma_x+i \sigma_y)^{\otimes N}+(\sigma_x-i \sigma_y)^{\otimes N}\right]
\end{equation}
which is equivalent to the general correlation when $f=1/2$. It means the coefficient of probability weighting function becomes,
\begin{equation}
g_{\vec{\alpha},\vec{m}}^M=(-1)^{\sum_j\alpha_j}\mbox{Re}[i^{(\sum_j m_j)}]
\end{equation}
and it can be used to obtain the local realistic bound. The bounds will be
\begin{eqnarray}
\max_{\vec{\alpha}}\left[\sum_{\vec{m}}g_{\vec{\alpha},\vec{m}}^M\right]&=&2^{(N-1)/2}~~~ \mbox{for odd} ~N, \nonumber\\
&=&2^{N/2}~~~ \mbox{for even} ~N.
\end{eqnarray}
With the coefficient $g_{\vec{\alpha},\vec{m}}^M$, the correlation in terms of probability distribution can be obtained as
\begin{eqnarray}
\langle M\rangle&=&\sum_{\vec{\alpha}}\sum_{\vec{m}} g^M_{\vec{\alpha},\vec{m}} p(\vec{\alpha}|\vec{m})\\
&=&\sum_{\vec{m}\in\mbox{all}}\mbox{Re}[i^{(\sum_j m_j)}]\\
&&~\times\Big[p\left(\mbox{even \# up}|\vec{m}\right)-p\left(\mbox{odd \# up}|\vec{m}\right)\Big]\nonumber
\end{eqnarray}
whose local realistic bound is violated by quantum state at a large scale as it can be found in the original work \cite{Mermin90}.

As it is discussed in \cite{Zukowski02}, the most general Bell function in two binary outcome measurements at $N$ sites can be obtained from the generating function of all the correlations,
\begin{equation}
\sum_{s_1,\cdots,s_N=\pm1} S(s_1,\cdots,s_N)\prod^N_{j=1}[A_j(0)+s_j A_j(1)]=\pm2^N
\end{equation}
where $S(s_1,\cdots,s_N)$ stands for an arbitrary function of the summation indices $s_1,\cdots,s_N\in\{-1,1\}$, such that their values are only $\pm 1$, {\it i.e.}, $S(s_1,\cdots,s_N)=\pm 1$. Since a general correlation function is defined as $E(m_1,m_2,\cdots)=\langle\prod_j A_j(m_j)\rangle$, the constrtaint for the convex sum of correlation functions are given
\begin{eqnarray}
\label{eq:ZB1}
&&\Big|\sum_{s_1, \cdots, s_N=\pm 1} S(s_1,\cdots,s_N)\\
&&~~~~~\times\sum_{\{m_j\}=0}^1s_1^{m_1}\cdots s_N^{m_N}E(m_1, \cdots, m_N)\Big|
\le 2^N.
\nonumber
\end{eqnarray}
Furthermore, compliance of the constraints for arbitrary choices of $S(s_1,\cdots,s_N)\in\{-1,1\}$ can be equated with a condition for a single correlation function. The correlation function by ZB is
\begin{eqnarray}
ZB&=&\sum_{s_1, \cdots, s_N=\pm 1}\\
&&~~\times\Big|\sum_{\{m_j\}=0}^1s_1^{m_1}\cdots s_N^{m_N}E(m_1,\cdots, m_N)\Big|\le 2^N\nonumber
\end{eqnarray}
whose validation gurantees the satisfaction of inequalities (\ref{eq:ZB1}) for any choice of $S(s_1, s_2,\cdots)$.

Comparing ZB function to the generic Bell function, the sum of all the correlation functions $E(m_1,m_2, \cdots)$ can be obtained when $f(\vec{n}_c)=f(c_1 n_1, c_2 n_2,\cdots)$ are specified as $f^{ZB}(c_1, c_2,\cdots)=(1-i)/2$. Under the circumstance, the generic Bell function becomes
\begin{equation}
G_{N,2,2}^{ZB, c}=\sum_{\{m_j\}=0}^1 P(m_1,m_2,\cdots)  c_1^{m_1} c_2^{m_2}\cdots E(m_1, m_2, \cdots)
\end{equation}
where the parity function $P(m_1,m_2,\cdots)\in\{-1,1\}$ takes its value $1$ for $[(\sum_j m_j) \mod 4] \in\{0, 1\}$ and $-1$ for $[(\sum_j m_j) \mod 4] \in\{2,  3\}$. All the Bell function in this setting can be found from the convex sum of the $G$ functions and it reads
\begin{equation}
ZB=\sum_c \left|G^{ZB, c}_{N,2,2}\right|\le 2^N
\end{equation}
which constitute the most general non-locality criteria in the given setting.

\subsection{Optimization for local realistic bound}
In the prvious section, it has been shown that the local realistic bound is found as $B_{LR}=\max_{\alpha}\left[\sum_{\vec{m}}g_{\vec{\alpha},\vec{m}}\right]$. From the relationship between $g_{\vec{\alpha}, \vec{m}}$ and $f(\vec{n})$, the bound can be expressed in terms of correlation weighting factor $f(\vec{n})$ as
\begin{eqnarray}
B_{LR}&=& \max_{\alpha}\left[\sum_{\vec{m}}g_{\vec{\alpha},\vec{m}}^c\right]\\
&=&\max_{\vec{\alpha}}\left[\sum_{\vec{n}} f(\vec{n})\prod_{j=1}^N\left(\sum_{m_j}\omega^{c_j n_j [\alpha_j(m_j)+\frac{m_j}{k}]}\right)\right]+c.c.\nonumber\\
&=&\max_{\vec{\alpha}}\sum_{\vec{n},\vec{m}} 2|f(\vec{n})|\cos\left[n \theta_f +\frac{2\pi\vec{n}_c}{d}\cdot\left(\vec{\alpha}+\frac{\vec{m}}{k}\right)\right].\nonumber
\label{eq:bound}
\end{eqnarray}
In order to make the local realistic bound optimal, the upper bound of $B_{LR}$ is evaluated after an appropriate parameterization $\vec{\alpha}=(\alpha_1(m_1),\alpha_2(m_2),\cdots)$. In the following, we show how to derive the local realistic bound for an arbitrary choice of $f(\vec{n})$ in a simple case.

In general, it is known that the function for the local realistic bound is not possible to be evaluated trivially. It is mainly because the maximzation of the function with respect to the measurement values is usually not straightforward and it is possible through the optimal specification of $kN$ independent parameters. Furthermore, it can be proved that the number can be reduced into $kN+1-N$ due to the symmetry in the trigonomateric function. When the system become simple, the straightforward maximization can be obtained.

For example, when the systems are in a simple case, $(2,k,2)$, the function takes the form
\begin{eqnarray}
B^{2,k,2}_{LR}&=&\max_{\alpha,\beta}\Big\{ 2 |f(1)| \sum_{m_1,m_2} \cos\left[\pi \alpha_{m_1}+\pi\beta_{m_2}\right]\nonumber\\
&&~~~~~~~~~~~~\times\cos\left[\theta_f+\frac{\pi(m_1+m_2)}{k}\right]\Big\}
\end{eqnarray}
and the optimization can be made through the specification of $\alpha_j(m_j)$. Through the assignement of the values $\alpha_{m_1}$ and $\beta_{m_2}$, the parity values of $\cos$ terms will be determined. Explicitly, the function can be expended
\begin{widetext}
\begin{eqnarray}
B^{2,k,2}_{LR}&=&2 |f(1)| \max_{\vec{\alpha}}\Big\{
\cos(\theta_f)
\left[(-1)^{\alpha_0+\beta_0}-(-1)^{\alpha_1+\beta_{k-1}}-(-1)^{\alpha_2+\beta_{k-2}}
\cdots-(-1)^{\alpha_{k-1}+\beta_1}
\right]\nonumber\\
&&~~~~~~~~~~~~~~~+\cos\left(\theta_f+\frac{\pi}{k}\right)[(-1)^{\alpha_0+\beta_1}+(-1)^{\alpha_1+\beta_0}-(-1)^{\alpha_2+\beta_{k-1}}
\cdots-(-1)^{\alpha_{k-1}+\beta_2}]\\
&&~~~~~~~~~~~~~~~+\cdots+\cdots~~~~~~~~~~~~~~~~~~~~~~~~~~~~\Big\}\nonumber\\
&=&2 |f(1)|\Big\{ k \cos(\theta_f)+ (k-2)\cos\left(\theta_f+\frac{\pi}{k}\right)+\cdots\cdots\Big\}
=2 |f(1)|\sum_{l=0}^{k-1} (k-2l )\cos\left(\theta_f+\frac{\pi l}{k}\right)\nonumber
\end{eqnarray}
\end{widetext}
where the maximization can be attained in the range $0\le\theta_f\le\frac{\pi}{2k}$. In the optimization, the parameters are specified in order to make the coefficient of the larger cosin term weighted more by assignment $\alpha_0=\alpha_1=\alpha_2=\cdots=\alpha_{k-1}=0$, $\beta_0=0$ and $\beta_1=\beta_2=\cdots=\beta_{k-1}=1$. It can be proved that the value of function $B^{2,k,2}_{LR}$ is maximum and the same procedure can be applied to the other range of $\theta_f$. In the other value of $\theta_f$, it also can be prove that the same maximum can be obtained. The result provides the local realistic bound for the setting $(2,k,2)$ which is the recent Bell test setting given by Epping {\it et. al.} \cite{Epping13}. Another general class of local realistic bound, (N,2,d)-class, has also been analyized in \cite{Bae14} for the specific choice of $f(n)$, as $|f(n)|=1$ and $\theta_f=\pi/4$.


\begin{thebibliography}{}
\bibitem{Bell64} Bell J S, \emph{Physics} {\bf 1}, 195 (1964).
\bibitem{Clauser69} J.F. Clauser, M.A. Horne, A. Shimony and R.A. Holt, \emph{Phys. Rev. Lett.} {\bf 23}, 880 (1969); N.D. Mermin, \emph{Phys. Rev. D} {\bf 22}, 356 (1980);G.  Svetlichny, \emph{Phys. Rev. D} {\bf 35}, 3066 (1987); M. Ardehali, \emph{Phys. Rev. A} \textbf{46}, 5375 (1992); N. Gisin and A. Peres, \emph{Phys. lett. A} {\bf 162}, 15 (1992); A.V. Belinskii and N. D. Klyshko, \emph{Phys. Usp.} \textbf{36}, 653 (1993);N. Gisin and H. Bechmann-Pasquinucci, \emph{Phys. Lett. A} \textbf{246}, 1 (1998); I. Pitowsky and K. Svozil, \emph{Phys. Rev. A} {\bf 64}, 014102 (2001);C. Sliwa, \emph{Phys. Lett. A} {\bf 317}, 165 (2003);D. Collins and N. Gisin, \emph{J. Phys. A: Math Gen.} {\bf 37}, 1775 (2004); S.W. Lee and D. Jaksch, \emph{Phys. Rev. A} {\bf 80}, 010103(R) (2009);J. Bancal, C. Branciard, N. Brunner, N. Gisin and Y. Liang, \emph{J. Phys. A: Math. Theor.} {\bf 45}, 125301 (2012);A. Tavakoli, S. Zohren and M. Pawlowski, \emph{J. Phys. A: Math. Theor.} {\bf 49}, 14 (2016).
\bibitem{Mermin90} N.D. Mermin, \emph{Phys. Rev. Lett.} {\bf 65}, 1838 (1990).
\bibitem{Peres99} A. Peres, \emph{Foundations of Physics} {\bf 29}, 589 (1999).
\bibitem{Werner01a} R. F. Werner and M. M. Wolf, \emph{Phys. Rev. A} {\bf 64}, 032112 (2001).
\bibitem{Zukowski02} M. Zukowski and C. Brukner, \emph{Phys. Rev. Lett.} \textbf{88}, 210401(2002).
\bibitem{Seevinck02} M. Seevinck and G. Svetlichny, \emph{Phys. Rev. Lett.} \textbf{89}, 060401 (2002).
\bibitem{Collins02} D. Collins, N. Gisin, N. Linden, S. Massar and S. Popescu, \emph{Phys. Rev. Lett.} {\bf 88}, 040404 (2002);S. Zohren and R. D. Gill, \emph{Phys. Rev. Lett.} {\bf 100},120406 (2008).
\bibitem{Barrett06} J. Barrett, A. Kent and S. Pironio, \emph{Phys. Rev. Lett.} {\bf 97}, 170409 (2006).
\bibitem{Son06} W. Son, J. Lee and M.S. Kim, \emph{Phys. Rev. Lett.} {\bf 96}, 060406 (2006); W. Son, C. Brukner and M. S. Kim, \emph{Phys. Rev. Lett.} {\bf 97}, 110401 (2006).
\bibitem{Deng09} D. L. Deng, Z. S. Zhou and J. L. Chen, \emph{Phys. Rev. A} {\bf 80}, 022109 (2009); J. L. Chen, D. L. Deng, H. Y. Su, C. Wu and C. H. Oh, \emph{Phys. Rev. A} {\bf 83}, 022316 (2011); J. D. Bancal, N. Brunner, N. Gisin and Y. C. Liang, \emph{Phys. Rev. Lett.} \textbf{106}, 020405 (2011);A. Cabello, \emph{Phys. Rev. Lett.} {\bf 114}, 220402 (2015).
\bibitem{Grandjean12}B. Grandjean, Y.C. Liang, J. D. Bancal, N. Brunner and N. Gisin, \emph{Phys. Rev. A} \textbf{85}, 052113 (2012).
\bibitem{Epping13} M. Epping, H. Kampermann and D. Bruss, \emph{Phys. Rev. Lett.} {\bf 111}, 240404 (2013).
\bibitem{Cabello14}A. Cabello, S. Severini, and A. Winter, \emph{Phys. Rev. Lett.} {\bf 112}, 040401 (2014).
\bibitem{Werner01} R. F. Werner and M. M. Wolf, \emph{Quantum information \& computation} {\bf 1} No.3, (2001).
\bibitem{Vertesi14} T. V\'ertesi and N. Brunner, \emph{Nature Communications} {\bf 5}, 5297 (2014).
\bibitem{Tura14} J. Tura, R. Augusiak, A. B. Sainz, T. Vertesi, M. Lewenstein and A. Acin Science {\bf 344}, 1256 (2014).
\bibitem{Laskowski04} W. Laskowski, T. Paterek, M. Zukowski and C. Brukner, \emph{Phys. Rev. Lett.} {\bf 93}, 200401 (2004).
\bibitem{Bancal10} J. Bancal, N. Gisin and S. Pironio, J. Phys. A: Math. Theor. {\bf 43}, 385303 (2010).
\bibitem{Schrodinger35}E. Schrodinger, \emph{Naturwissenschaften} {\bf 23}, 800, 823, 844 (1935); translation in Quantum Theory and Measurement, J. A. Wheeler and W. H. Zurek, eds. (Princeton University Press, Princeton, NJ, 1983) , p. 152.
\bibitem{Arnault12} F. Arnault, J. Phys. A: Math. Theor. {\bf 45} 255304 (2012).
\bibitem{Rockafellar70} R. T. Rockafellar, Convex Analysis (Princeton University Press, Princeton, NJ), p. 200 (1970).
\bibitem{Bae14} G. Bae and W. Son,  Curr. Appl. Phys. {\bf 16} 378 (2016).
\bibitem{Sup} See the appendix section.
\bibitem{Masanes02} L. Masanes, Quantum Inf. Comput. {\bf 3} 345 (2002).
\end{thebibliography}
\end{document}